\documentclass{PoS}

\title{Lattice QCD calculation of isospin breaking effects due to the up-down mass
difference}

\ShortTitle{Lattice QCD calculation of strong isospin breaking effects}

\author{\speaker{Francesco Sanfilippo}\\
  ~Laboratoire de Physique Th\'eorique (B\^at.~210)~\footnote{Laboratoire de Physique Th\'eorique est une unit\'e mixte de recherche du CNRS, UMR 8627.}\\
  Universit\'e Paris Sud, F-91405 Orsay-Cedex, France\\
  and INFN, Sez. di Roma, Piazzale Aldo Moro 5, I-00185 Roma, Italy
  E-mail: \email{francesco.sanfilippo@th.u-psud.fr}}

\author{G.M.~de Divitiis, R.~Frezzotti, R.~Petronzio, G.C.~Rossi, N.~Tantalo\\
  Dip. di Fisica, Universit\`a di Roma ``Tor Vergata" and INFN,
  Via della Ricerca Scientifica 1, 00133 Rome, Italy}

\author{P.~Dimopoulos\\
  Dip. di Fisica, Universit\`a di Roma ``Tor Vergata"
  Via della Ricerca Scientifica 1, 00133 Rome, Italy}

\author{V.~Lubicz, C.~Tarantino\\
  Dip. di Fisica, Universit{\`a} Roma Tre and INFN,
  Via della Vasca Navale 84, 00146 Rome, Italy}

\author{G.~Martinelli\\
  SISSA, Via Bonomea 265, 34136, Trieste, Italy\\
  and INFN, Sez. di Roma, Piazzale Aldo Moro 5, I-00185 Roma, Italy}

\author{S.~Simula\\
  INFN, Sez. di Roma Tre, Via della Vasca Navale 84, 00146 Rome, Italy}

\abstract{We present a new method to evaluate with high precision the
  isospin breaking effects due to the mass difference between
  the up and down quarks using lattice QCD. Our proposal is applicable
  in principle to any hadronic observable which can be computed on the
  lattice. It is based on the expansion of the path-integral in powers
  of the small parameter $m_d - m_u$. In this talk we discuss how to apply this
  method to compute the leading isospin breaking effects for several
  physical quantities of interest: the kaon masses, the kaon
  decay constants and the neutron-proton mass splitting.}

\FullConference{XXIX International Symposium on Lattice Field Theory \\
                 July 10 - 16 2011\\
                 Squaw Valley, Lake Tahoe, California}

\usepackage{graphics,epsfig,units}
\usepackage{xspace}
\usepackage{amsfonts,amsmath,amssymb} 
\usepackage{cite}

\newcommand{\bea}{\begin{eqnarray}}
\newcommand{\eea}{\end{eqnarray}}
\newcommand{\beq}{\begin{equation}}
\newcommand{\eeq}{\end{equation}}

\def\slash#1{\mbox{$\not\!\! #1$}}

\def\simge{\mathrel{\rlap{\raise 0.511ex \hbox{$>$}}{\lower 0.511ex
 \hbox{$\sim$}}}}
\def\simle{\mathrel{\rlap{\raise 0.511ex \hbox{$<$}}{\lower 0.511ex
 \hbox{$\sim$}}}}
\def\slash#1{\setbox0=\hbox{$#1$}\dimen0=\wd0 \setbox1=\hbox{/} \dimen1=\wd1
 \ifdim\dimen0>\dimen1 \rlap{\hbox to \dimen0{\hfil/\hfil}} #1
 \else \rlap{\hbox to \dimen1{\hfil$#1$\hfil}} / \fi}

\newcommand{\sla}[1]%
        {\kern .25em\raise.18ex\hbox{$/$}\kern-.6em #1}

\newcommand{\gou}{\raisebox{-0.4\totalheight}{\includegraphics[scale=.4]{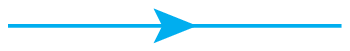}}}
\newcommand{\god}{\raisebox{-0.4\totalheight}{\includegraphics[scale=.4]{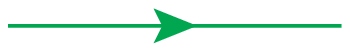}}}
\newcommand{\gol}{\raisebox{-0.4\totalheight}{\includegraphics[scale=.4]{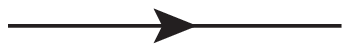}}}
\newcommand{\goi}{\raisebox{-0.4\totalheight}{\includegraphics[scale=.4]{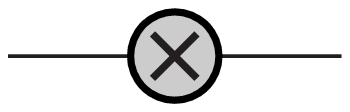}}}

\newcommand{\gdsi}{\raisebox{-0.4\totalheight}{\includegraphics[scale=.3]{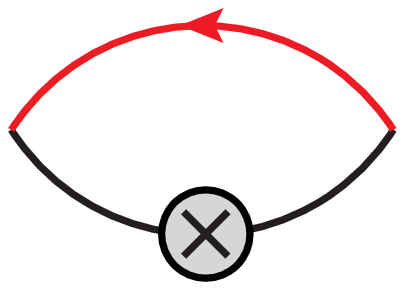}}}

\newcommand{\gdsu}{\raisebox{-0.4\totalheight}{\includegraphics[scale=.3]{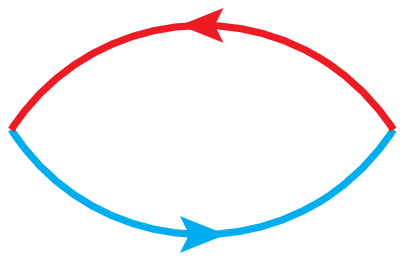}}}

\newcommand{\gdsd}{\raisebox{-0.4\totalheight}{\includegraphics[scale=.3]{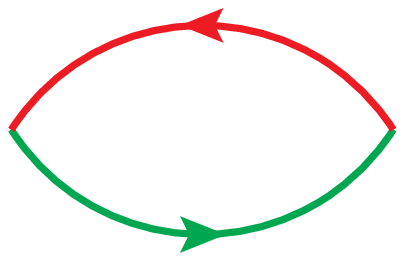}}}
\newcommand{\gdsl}{\raisebox{-0.4\totalheight}{\includegraphics[scale=.3]{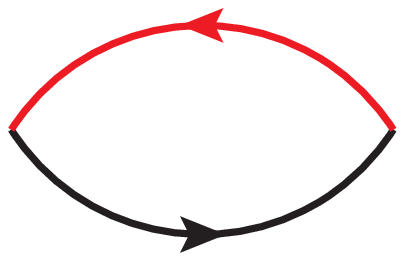}}}

\newcommand{\ins}{\raisebox{-0.4\totalheight}{\includegraphics[scale=.3]{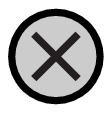}}}

\newcommand{\bear}[1]{\begin{equation}\begin{array}{#1}}
\newcommand{\eear}{ \end{array}\end{equation}}
\newcommand{\barr}[1]{\begin{array}{#1}}
\newcommand{\earr}{\end{array}}

\usepackage[title,titletoc]{appendix}

\renewcommand{\thetable}{\Alph{table}}

\begin{document}

\section{Introduction}
Nowadays, with the increasing precision of the experimental
determinations of many physical quantities, and in some cases with the
improvement of the theoretical predictions, the control over isospin
breaking effects is becoming phenomenologically relevant.

In the past, isospin breaking effects due to the light quarks mass
difference have been accommodated within the chiral
perturbation theory (ChPT) framework
\cite{Gasser:1984gg,Gasser:1984pr,Amoros:2001cp,Bijnens:2007xa,
  Kastner:2008ch,Cirigliano:2011tm}, while several attempts to compute
electromagnetic effects for the hadron spectroscopy in lattice QCD
have been presented~\cite{Duncan:1996xy,Basak:2008na,
  Blum:2010ym,Portelli:2010yn}.

Isospin breaking effects in the Standard Model are induced both from light
quark masses (QCD effects) and charges difference (QED effects). In the
real world the separation between the two terms is conventional due to different
additive renormalization of $u,d$ quark masses in presence of electromagnetism.
In the present work we consider the well-defined theoretical limit in which 
the electromagnetic interactions are switched off, and present
 a new method to compute the leading QCD isospin breaking
effects with high precision. The method is based on the expansion of
the lattice path-integral in powers of the small parameter $m_d - m_u$
and is applicable in principle to any hadronic observable which can be
computed on the lattice.

Till now we have applied it
to the computation of the kaon masses, the kaon decay constants,
and the neutron-proton mass splitting. In the
future we plan to apply the method to other physical quantities, to
include QED corrections and to try also the calculation of
next-to-leading corrections such as the $\pi^+$-$\pi^0$ mass
difference.

The results presented in this talk have been published in \cite{deDivitiis:2011eh}.
Here we present a shorter version of the published work: we marked in the
text the parts which are discussed in more details in the paper.

\section{Description of the method}
\label{sec:method}
Let us start by considering the evaluation
of a generic euclidean correlation function $\langle {\cal O}\rangle $
used to extract information about physical quantities as masses, decay
constants, form factors etc.,
\bea 
\langle {\cal O}\rangle = 
\frac{1}{\mathcal{Z}} \int{ D\phi \ {\cal O} \, e^{-S}}\quad,\quad\quad \mathcal{Z}={\int{ D\phi \ e^{-S} }}\, .
\label{eq:fi}
\eea 
We can split the Lagrangian into $SU(2)_V$ symmetric and isosping violating terms,
\bea
  {\cal L}  = {\cal L}_{0} - \Delta m_{ud} \hat {\cal L} = {\cal L}_{kin}  + m_{ud}\bar q q - \Delta m_{ud} \bar q \tau^3 q  , 
    \label{eq:lag} 
\eea
where $q^T=(u,d)$, $m_{ud}=(m_d+m_u)/2$ and $\Delta m_{ud}=(m_d-m_u)/2$.  By
expanding at first order the exponential of the action, $S=\sum_x
{\cal L}(x)$, with respect to $\Delta m_{ud}$ we obtain:
\bea
\langle {\cal O}\rangle 
\simeq
\frac{\int{ D\phi \ {\cal O}\, (1+  \Delta m_{ud} \, \hat S)\, e^{-S_0} }}
{\int{ D\phi \ \, (1+  \Delta m_{ud} \, \hat S) \, e^{-S_0} }}
= 
\frac{\langle {\cal O}\rangle_0 +  \Delta m_{ud} \, \langle {\cal O}\hat S\rangle_0 }
{1+ \Delta m_{ud} \, \langle \hat S\rangle_0 }\, 
=\langle {\cal O}\rangle_0 +  \Delta m_{ud} \, \langle {\cal O}\hat S\rangle_0 \, ,
\eea
where $\langle \cdot \rangle_0$ represent the vacuum expectation value
in the isospin symmetric theory and $\hat S$ is the isospin breaking
term $\hat S=\sum_x{[\bar q \tau_3 q](x)}=\sum_x{[\bar u u - \bar d d](x)} $.
The correction in the denominator vanishes, $\langle \hat
S\rangle_0=0$, because of isospin symmetry.
We can now describe a general recipe to be used in order to
compute leading QCD isospin breaking effects on the lattice:

\begin{itemize}
\item consider a given correlation function in the full theory, i.e. with $m_u\neq m_d$, and for each gauge configuration draw all the fermionic Wick contractions;
\item expand the up and down quark propagators with respect to $\Delta m_{ud}$ according to
\bea
G_{u/d}(x_1,x_2) = G_\ell(x_1,x_2) \pm \Delta m_{ud} \ \sum_y{G_\ell(x_1,y)\ G_\ell(y,x_2)}+\cdots\, ;
\label{eq:propsexpformulae}
\eea
\item retain the terms linear in $\Delta m_{ud}$ and compute the corresponding diagrams.
\end{itemize}

Eq.~(\ref{eq:propsexpformulae}) can be represented diagrammatically
as follows
\bea
\begin{array}{c}
\overset{u} {\gou} \\
\overset{d} {\god} \\
\end{array}
 & = \gol \pm \goi  + \cdots \, ,
\eea
where plus holds for up (light blue) and minus for down (green) quarks.
Black lines from $x$ to $y$ refer to $G_\ell(x-y)= \langle \ell(x)\bar \ell(y) \rangle $,
 the propagator with the symmetric
mass $m_{ud}$ in the isospin symmetric theory, whereas the cross represents the
insertion of the renormalization group invariant quantity
$\ins = \Delta m_{ud} \,  \sum_z{\bar \ell(z) \ell (z)}$.

Note that the square of the $G_\ell$ propagator entering
eq.~(\ref{eq:propsexpformulae}) can be easily calculated on the
lattice by using $G_\ell$ itself as the source vector of a new
inversion.

\section{Kaon masses and decay constants}
\label{sec:kaontwopoint}
In this section we discuss in detail the strategy used to derive the isospin corrections to the kaon masses and decay constants.
 To this end we start by expanding the Euclidean two-point pseudoscalar correlation functions of kaons around the isospin symmetric point:
\bea
C_{K^+K^-}(t) = \sum_{\vec x}
e^{-i\vec{p}\cdot \vec{x}}
\langle\, \bar u \gamma_5 s(x)\ \bar s \gamma_5 u(0)\, \rangle =
-\overset{s} {\underset{u} {\gdsu}}=   -\gdsl -   \gdsi + {\cal O}(\Delta m_{ud})^2\, ,
\nonumber \\
\nonumber \\
C_{K^0K^0}(t) = \sum_{\vec x}
e^{-i\vec{p}\cdot \vec{x}}
\langle\, 
\bar d \gamma_5 s(x)\ 
\bar s \gamma_5 d(0)
\, \rangle \, =
-\overset{s} {\underset{d} {\gdsd}}= -  \gdsl +   \gdsi + {\cal O}(\Delta m_{ud})^2  \, .   
\label{eq:kpcorr}
\eea
The spectral decomposition of $C_{K^0K^0}$ (the analysis of $C_{K^+K^-}$ proceeds along similar lines) is
\bea 
C_{K^0K^0}(\vec p, t) &=& 
\sum_{\vec x} 
e^{-i \vec p \cdot \vec x } 
\langle \bar d \gamma_5 s(\vec x,t)\, \bar s \gamma_5 d(0)\rangle \, 
= \frac{G_{K^0}^2}{2E_{K^0}}\ e^{-E_{K^0}t}+\cdots\, ,
\label{eq:spectral}
\eea
By differentiating eq.~(\ref{eq:spectral}) with respect to $\Delta m$ it is easy to see that from the ratio of the two correlators
\bea
\delta C_{KK}(\vec p, t) 
=\frac{\Delta C_{KK}(\vec p, t)}{C_{KK}(\vec p, t)}
=-\frac{\gdsi}{\gdsl}
=
\delta\left(\frac{G_K^2}{2E_K}\right)-t\Delta E_K
+\cdots \, ,
\label{eq:ckkratiocont}
\eea
it is possible to extract the leading QCD isospin breaking corrections to kaon energies and decay constants. Indeed $\Delta E_K$ appears
directly in the previous equation as the ``slope" with respect to $t$ whereas $\delta F_K$ can be extracted from the ``intercept" according to
\bea
F_K = (m_s+m_{ud})\frac{G_K}{M_K^2} \, ,\quad
\delta F_K = \frac{\Delta m_{ud}}{m_s+m_{ud}} + \delta G_K -2 \delta M_K \, .
\eea  
On a lattice of finite time extent $T$ with quark fields satisfying anti-periodic boundary conditions along the time direction and given our choice of the kaon source and sink operators,
the pseudoscalar densities, eq.~(\ref{eq:ckkratiocont}) has to be properly modified.

\begin{figure}[!t]
\begin{center}
\includegraphics[width=0.45\textwidth]{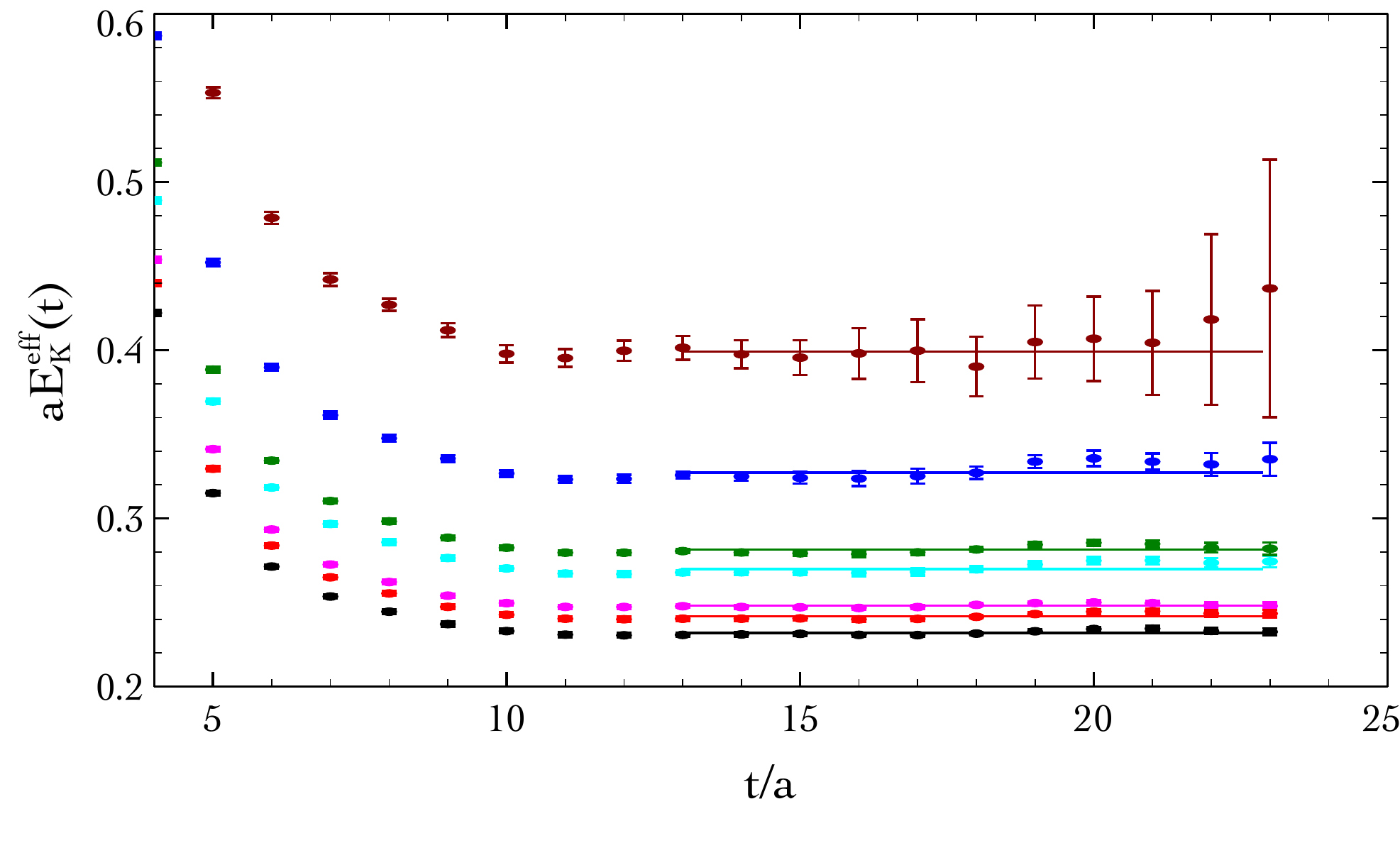}\hfill
\includegraphics[width=0.45\textwidth]{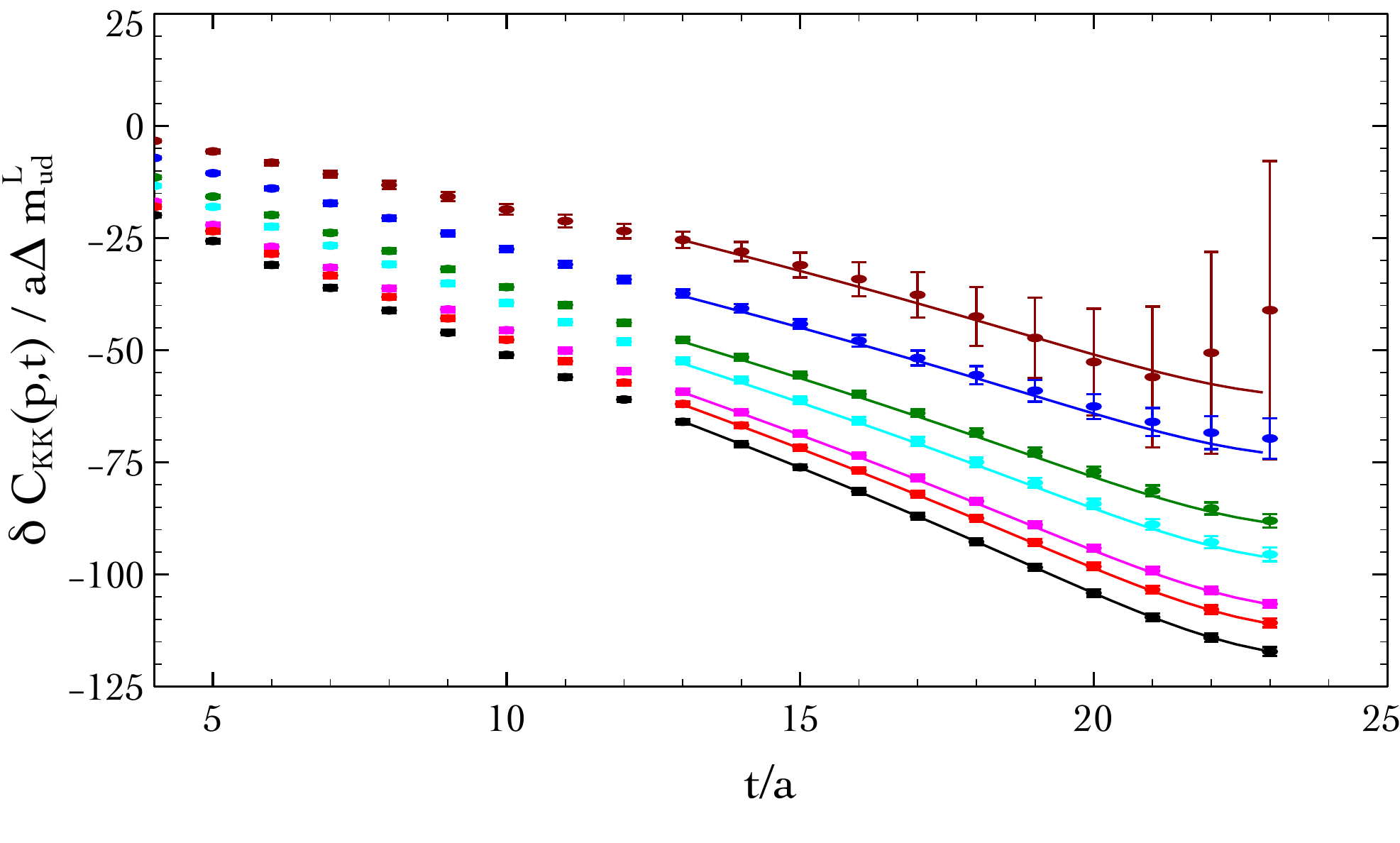}
\caption{
  \label{fig:Kslopes} \footnotesize
        {\it Left panel}: extraction of meson energies from the effective mass of $C_{KK}(\vec p,t)$.
        {\it Right panel}: fits of $\delta C_{KK}(\vec p,t) / a\Delta m_{ud}^L$: as it can be seen numerical data follow
        theoretical expectations. The data correspond to $\beta=3.9$, $am_{ud}^L=0.0064$, $am_{s}^L=0.0177$, with $am_{q}^L$ 
        the bare mass for quark $q$.
}
\end{center}
\vspace{-0.6cm}
\end{figure}

In this work we have used the $N_f=2$ dynamical gauge ensambles generated and made publicly available by the European Twisted Mass Collaboration.
These gauge configurations have been generated by using the so called Twisted Mass lattice discretization of the QCD action~\cite{Frezzotti:2000nk}. 
For the present analysis we have used the same 13 gauge ensembles taken into account in ~\cite{Blossier:2010cr}, using the same values of renormalization constants and lattice spacings
there mentioned, and considering a statistics of about 150 configurations per ensemble. The computation of all the correlation functions has been carried out on AURORA machine in Trento.
The full set of parameters can be found in the appendix A of \cite{deDivitiis:2011eh}, where we discuss also the relevance of isospin breaking effects at finite lattice spacing
caused by the Twisted Mass regularization.

As can be seen from Figure~\ref{fig:Kslopes}, $ \delta C_{KK}(\vec p, t)$ is determined with high precision, given the strong statistical correlation 
existing between the numerator and the denominator of the ratio in eq.~(\ref{eq:ckkratiocont}).

\begin{figure}[!t]
\begin{center}
\includegraphics[width=0.49\textwidth]{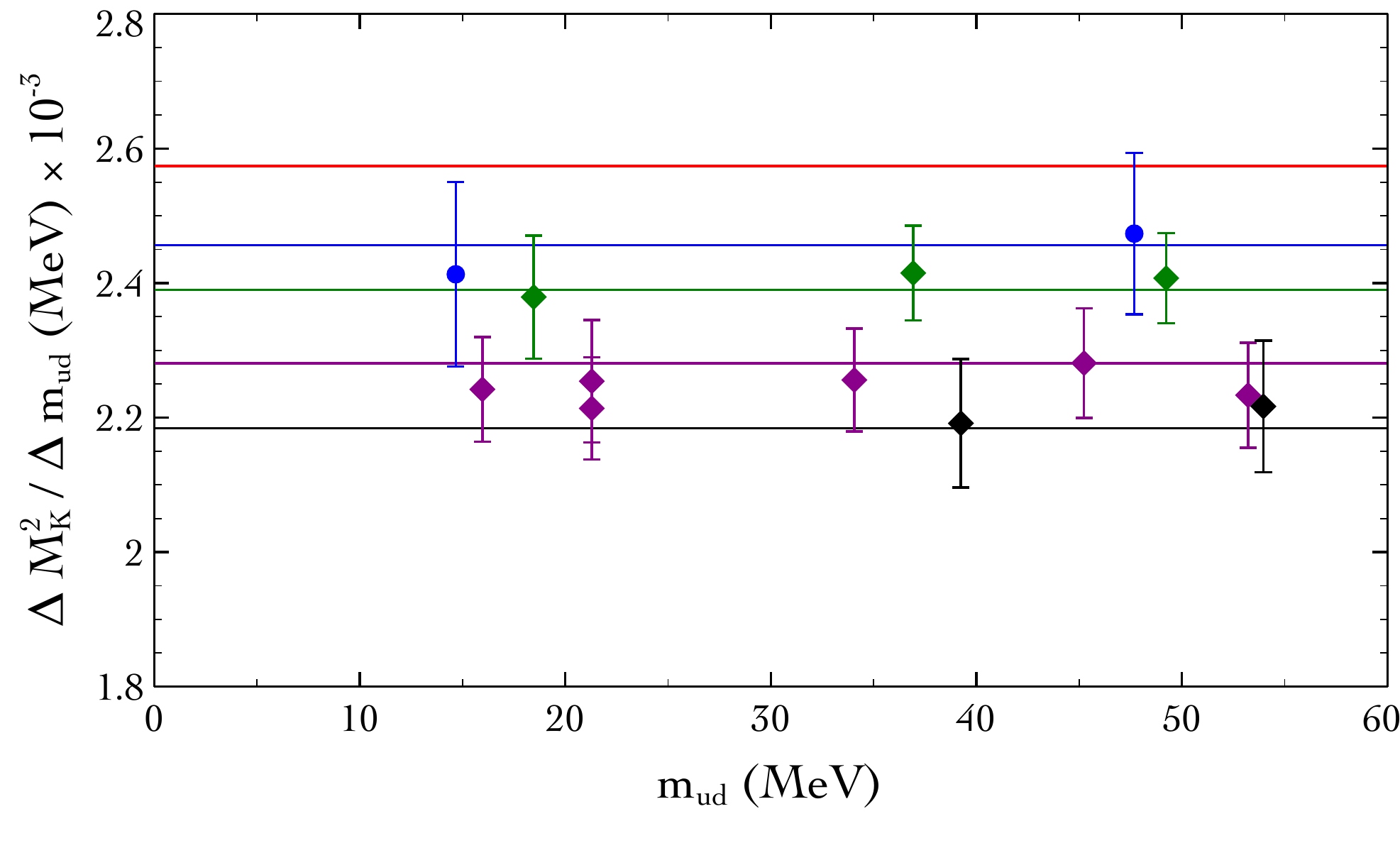}\hfill
\includegraphics[width=0.49\textwidth]{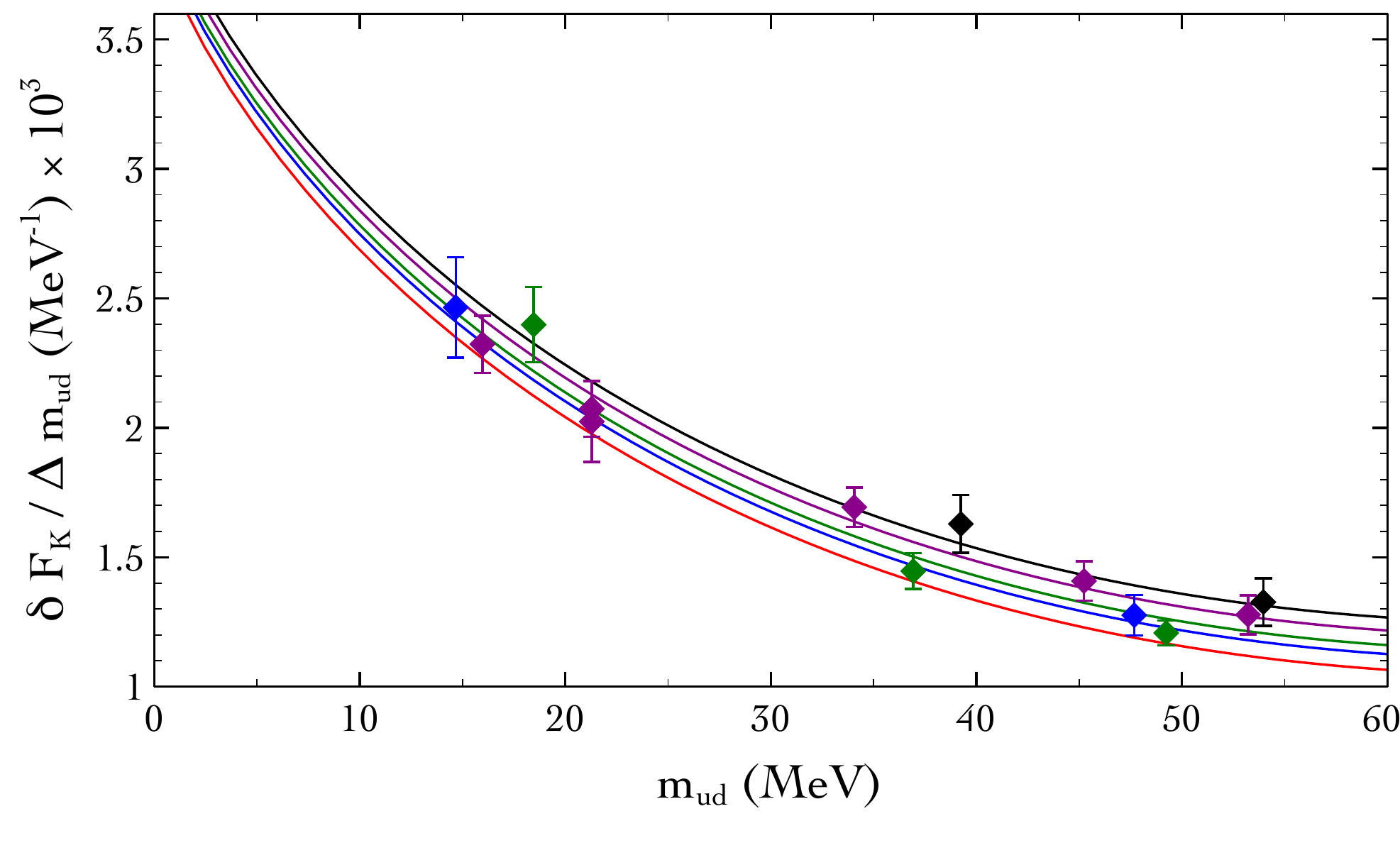}
\caption{
  \label{fig:Kchiral} \footnotesize
        {\it Left panel}: combined chiral and continuum extrapolations of $\Delta M_K^2/\Delta m_{ud}$, in physical units.
        {\it Right panel}: combined chiral and continuum extrapolations of $\delta F_K/\Delta m_{ud}$. Black points correspond to the coarser lattice spacing, $a=0.098$~fm, 
        dark magenta points correspond to $a=0.085$~fm, green points to $a=0.067$~fm and blue points to $a=0.054$~fm. Red lines are the results of the continuum extrapolations.}
\end{center}
\vspace{-0.6cm}
\end{figure}

Our results do not show a visible dependence with respect to $m_{ud}$ within the quoted errors. 
Therefore we simply extrapolated the correction to the meson mass square (which is a 
finite quantity in the chiral limit) to the continuum using the expression:
\bea
\left[\frac{\Delta M_K^2}{\Delta m_{ud}}\right](m_{ud},a)=
\left[\frac{\Delta M_K^2}{\Delta m_{ud}}\right]^{QCD}+C_M a^2 \, .
\label{eq:fitmk}
\eea
In the left panel of Figure~\ref{fig:Kchiral} we show the continuum extrapolation of $\Delta M_K^2/\Delta m_{ud}$. 
Adding a linear and a (chiral) logarithmic term according to the formulae obtained within the unitary $SU(3)_L\times SU(3)_R$ in ref.~\cite{Gasser:1984gg}
leads to compatible results for $[\Delta M_K^2/\Delta m_{ud}]^{QCD}$, with a $3\%$ difference which we considered as an estimate of systematic error,
adding it in quadrature to the lattice uncertainty.

In the case of $\delta F_K/\Delta m_{ud}$ (right panel of Figure~\ref{fig:Kchiral}), the dependence upon $m_{ud}$ is significant, and we have included in the fitting function the leading
and next-to-leading terms expanded in powers of $m_{ud}/m_s$ plus a lattice artifact term, i.e.
\bea
\left[\frac{\delta F_K}{\Delta m_{ud}}\right](m_{ud},a)=
\left[\frac{\delta F_K}{\Delta m_{ud}}\right]^{QCD}
+C_F a^2+B_1 \left(m_{ud}-m_{ud}^{QCD}\right) + B_2\, m_{ud}\log\left(\frac{m_{ud}}{m_{ud}^{QCD}}\right) \, .
\label{eq:fitfk}
\eea
The systematics associated to this extrapolation has been estimated by replacing the logarithmic term with a quadratic one, and it has been found of the order of $5\%$.

Using the value $m_{ud}^{ QCD}=m_{ud}^{ QCD}(\overline{ MS},{ 2GeV})=3.6(2)$ MeV from refs.~\cite{Colangelo:2010et,Blossier:2010cr} we obtain
\bea
\left[\frac{\Delta M_{K}^{2}}{\Delta m_{ud}}\right]_{\overline{MS},2GeV}^{QCD}=2.57(8)\times10^{3}\ \mbox{MeV}\,,\quad\quad
\left[\frac{\delta F_{K}}{\Delta m_{ud}}\right]_{\overline{MS},2GeV}^{QCD}=3.3(3)\times10^{-3}\ \mbox{MeV}^{-1}\,.
\label{eq:fitresults}
\eea

Having neglected the QED effects from our calculations, we cannot directly use the experimental determination of $M_{K^0}^2-M_{K^+}^2$ to extract $[m_d-m_u]^{QCD}$.
By taking the Chiral Perturbation Theory (ChPT) estimate of the electromagnetic corrections to $M_{K^0}^2-M_{K^+}^2$ we can obtain the
theoretical value of the kaons squared mass difference in absence of electromagnetic interactions
 ~\cite{Duncan:1996xy,Basak:2008na,Blum:2010ym,Portelli:2010yn},
ref.~\cite{Colangelo:2010et}, which reads:
\bea
\left[M_{K^0}^2-M_{K^+}^2\right]^{QCD}=6.05(63)\times 10^3\ \mbox{MeV}^2\, .
\label{eq:input}
\eea 

Using eq.~(\ref{eq:input}) with our numerical results given in eq.~(\ref{eq:fitresults}), we get
\vspace{-7 pt}
\bea
\left[m_d-m_u\right]^{QCD}(\overline{MS},2GeV)=
2.35(8)(24)\ \mbox{MeV} 
\quad\quad
\left[\frac{F_{K^0}-F_{K^+}}{F_{K}}\right]^{QCD}\hspace{-10pt}=0.0078(7)(4)
\label{eq:kaonphysres}
\eea 
where the first error comes from our calculation and combines in quadrature statistics and systematics while the second comes from the uncertainty on QED effects.
At first order in $\Delta m_{ud}$, due to the fact that pions don't get corrections and that $K^+$ and $K^0$ get opposite corrections, we have
\bea
\left[\frac{F_{K^+}/F_{\pi^+}}{F_{K}/F_{\pi}}-1\right]^{QCD} &=& -0.0039(3)(2)\,.
\eea
This is significatively higher than the estimate obtained in ref.~\cite{Cirigliano:2011tm} by using chiral perturbation theory, where the value -0.0022(6) was found.

\section{Nucleon masses}
\label{sec:nucleons}
Having determined $\Delta m_{ud}^{QCD}$,  we can now predict the QCD part of 
the difference between the masses of the neutron and of the proton. We consider for proton the correlation function
\bea
C_{pp}^\pm(t)=
\sum_{\vec x}
\langle\, 
\left [\epsilon_{abc}(\bar u_a C\gamma_5 \bar d^T_b ) 
\bar u_c  \frac{1\pm \gamma^0}{2}\right](x)\, 
\left [\epsilon_{def}\frac{1\pm \gamma^0}{2} u_d
(u^T_e C\gamma_5 d_f ) \right](0)
\, \rangle \, ,
\eea
\noindent
and similarly for neutron, with $u\leftrightarrow d$.
To increase statistic, we consider the combinations
\bea
C_{nn}(t)= C_{nn}^+(t)-C_{nn}^-(T-t) \, , \quad \quad
C_{pp}(t)= C_{pp}^+(t)-C_{pp}^-(T-t) \, . 
\label{eq:nucleonscorrs_iso}
\eea
Quark fields have been ``Gaussian smeared'' according to the values found in ref.~\cite{Alexandrou:2008tn}.
The correction to such correlation function with respect to the isospin symmetric theory can be found by expanding eq.~(\ref{eq:nucleonscorrs_iso}) and computing all the required contractions,
as already shown in previous section in the case of kaons.
\begin{figure}[!t]
\begin{center}
\includegraphics[width=0.49\textwidth]{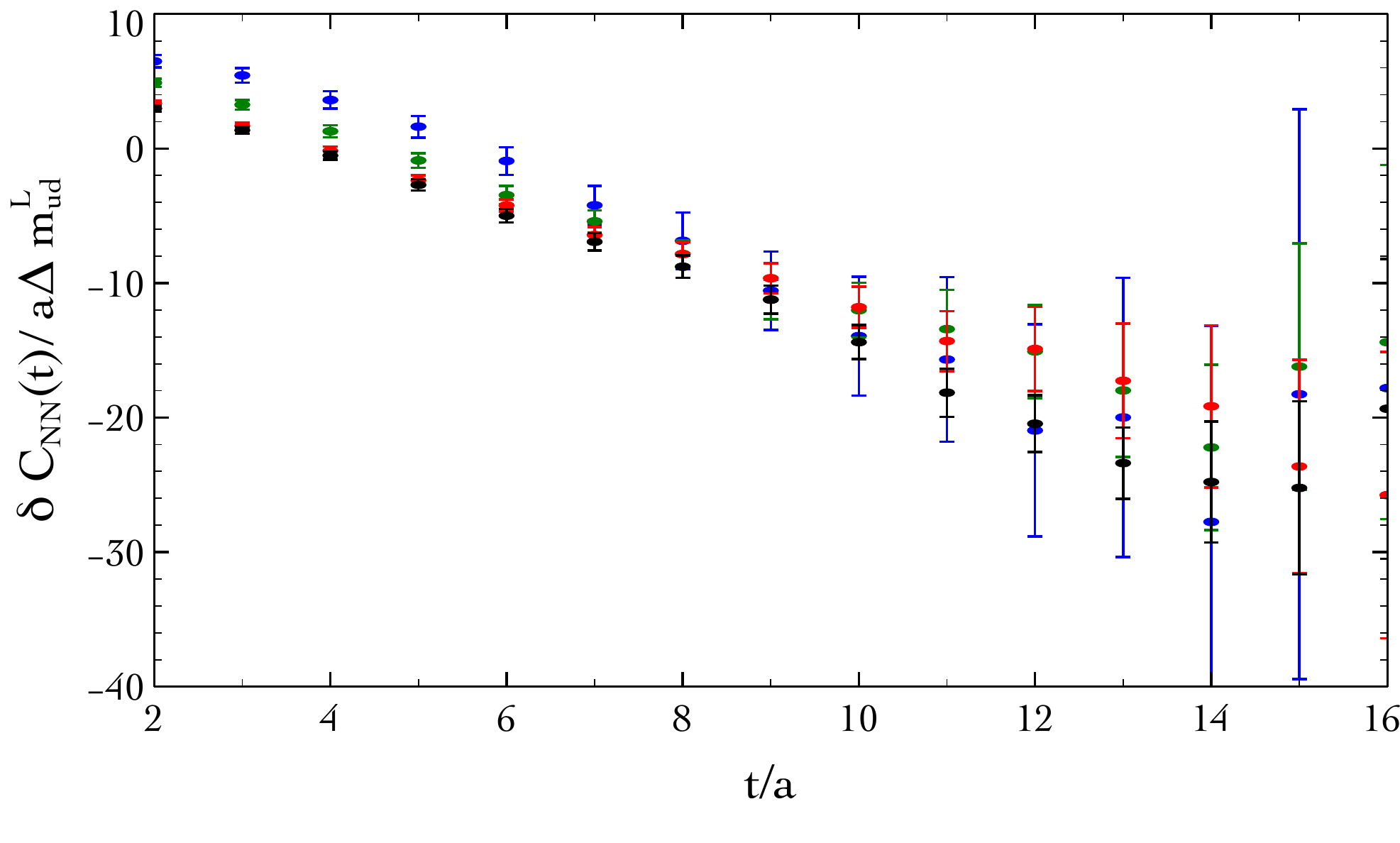}\hfill
\includegraphics[width=0.49\textwidth]{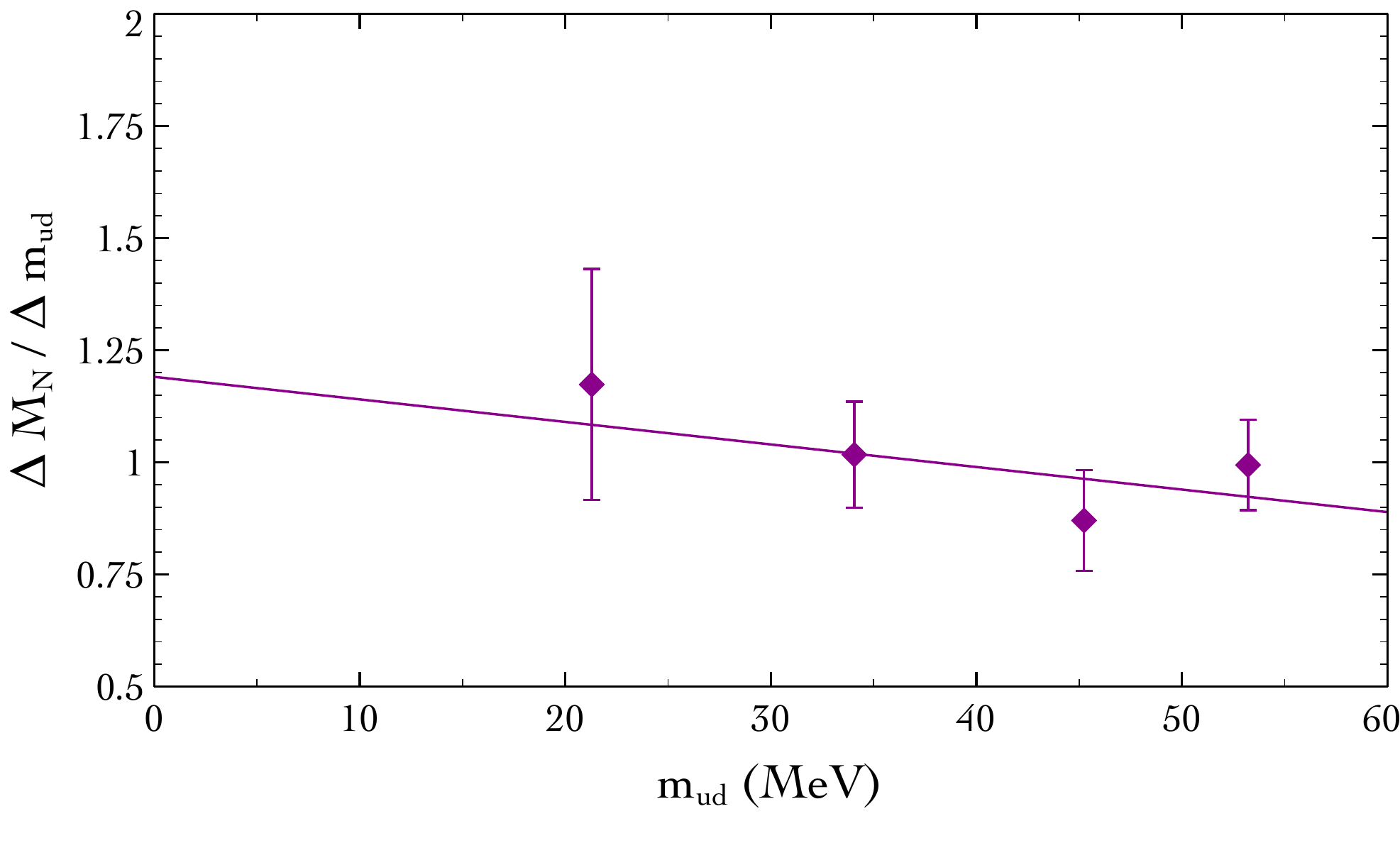}
\caption{\label{fig:nucleons} \footnotesize
  {\it Left panel}: Correlation functions $\delta C_{NN}(t)/a\Delta m_{ud}^L$.
  {\it Right panel}: Chiral extrapolation of $\Delta M_N/\Delta m_{ud}$.
  The data are at fixed lattice spacing $a=0.085$~fm for different values of $m_{ud}$.
}
\end{center}
\vspace{-0.6cm}
\end{figure}

The extraction of physical information from nucleon euclidean two point functions proceeds along the same lines described in detail in the case of the kaons.
By extracting the slope in  $t$ of $\delta C_{NN}(t)$,  we can determine $\Delta M_{N}=(M_n-M_p)/2$. In the left panel of Figure~\ref{fig:nucleons}
 we show  $\delta C_{NN}(t)/a\Delta m_{ud}^L$,  that we have fitted with the linear function: $\delta C_{NN}=c+t\Delta M_N$ .

In the right panel of Figure~\ref{fig:nucleons} we show the chiral extrapolation of $\Delta M_N/\Delta m_{ud}$ performed by using the following fitting function~\cite{Gasser:1982ap}
\bea
\left[\frac{\Delta M_N}{\Delta m_{ud}}\right](m_{ud})=
\left[\frac{\Delta M_N}{\Delta m_{ud}}\right]^{QCD}+B_N (m_{ud}-m_{ud}^{QCD})\, .
\eea
By using the results of the fit and the value of $\Delta m_{ud}^{QCD}$ given in eq.~(\ref{eq:kaonphysres}), we get
\bea
\left[M_n-M_p\right]^{QCD}= 2\Delta m_{ud}^{QCD} 
\left[\frac{\Delta M_N}{\Delta m_{ud}}\right]^{QCD}
=2.8(6)(3)\ \mbox{MeV}
\quad \times \quad 
\frac{\left[M_{K^0}^2-M_{K^+}^2\right]^{QCD}}{6.05\times 10^3\ \mbox{MeV}^2} \, ,
\eea 
where the first error takes into account the lattice uncertainties while the second comes from the uncertainty on QED contribution.
This is our best estimate at present but it has been obtained at fixed lattice spacing and with limited statistics. A refinement of this calculation is in progress.

\section{Conclusions and Outlooks}
\label{sec:outlooks}
In this talk we have proposed a new method to compute with high precision the QCD isospin breaking effects in relevant physical quantities at the lowest non trivial order in the up-down
mass difference. The method can be easily extended with minor modifications to higher orders. 
We have computed the corrections to meson and nucleon masses and meson decay constants, showing that, in spite of the limited statistics, our approach is already competitive,
or even better, than other calculations based on the effective QCD chiral lagrangian.

To obtain the complete physical results, our method has to be combined with the calculations of the electromagnetic corrections, which is currently under investigation. In this work, 
for a comparison with calculations in different theoretical frameworks, we have taken  the electromagnetic corrections to the meson masses  evaluated in ref.~\cite{Colangelo:2010et}.

As the method looks very promising, we are planning to extend this work to other physical observables, such as the form factors of semileptonic $K_{\ell3}$ decays, for which we
have presented preliminary results in the full paper \cite{deDivitiis:2011eh}.
\vspace{-0.3cm}
\section*{Acknowledgements}
Work partially supported  by the Programme IDEAS, ERC-2010-AdG, DaMESyFla
Grant Agreement Number: 267985, and by the MIUR (Italy) under the contracts PRIN08 and PRIN09. 
V.L. acknowledges the support of CNRS and the LPT, 
Universit\'e Paris-Sud 11, where part of this work was completed.
Computation performed on AURORA computing center.



\begin{thebibliography}{99}
\bibitem{Gasser:1984gg}
  J.~Gasser, H.~Leutwyler,
  Nucl.\ Phys.\  {\bf B250 } (1985)  465.

\bibitem{Gasser:1984pr}
  J.~Gasser, H.~Leutwyler,
  Nucl.\ Phys.\  {\bf B250 } (1985)  539.   

\bibitem{Amoros:2001cp}
  G.~Amoros, J.~Bijnens, P.~Talavera,
  Nucl.\ Phys.\  {\bf B602 } (2001)  87-108.
  [hep-ph/0101127].   
  
\bibitem{Bijnens:2007xa}
  J.~Bijnens, K.~Ghorbani,
  [arXiv:0711.0148 [hep-ph]].
  
\bibitem{Kastner:2008ch}
  A.~Kastner, H.~Neufeld,
  Eur.\ Phys.\ J.\  {\bf C57 } (2008)  541-556.
  [arXiv:0805.2222 [hep-ph]].  
  
\bibitem{Cirigliano:2011tm}
  V.~Cirigliano, H.~Neufeld,
  Phys.\ Lett.\  {\bf B700 } (2011)  7-10.
  [arXiv:1102.0563 [hep-ph]].

\bibitem{Duncan:1996xy}
  A.~Duncan, E.~Eichten, H.~Thacker,
  Phys.\ Rev.\ Lett.\  {\bf 76 } (1996)  3894-3897.
  [hep-lat/9602005].

\bibitem{Basak:2008na}
  S.~Basak {\it et al.} [ MILC Collaboration ],
  PoS {\bf LATTICE2008 } (2008)  127.
  [arXiv:0812.4486 [hep-lat]].
  
\bibitem{Blum:2010ym}
  T.~Blum {\it et al.},
  Phys.\ Rev.\  {\bf D82 } (2010)  094508.

\bibitem{Portelli:2010yn}
  A.~Portelli {\it et al.} [ Budapest-Marseille-Wuppertal Collaboration ],
  PoS {\bf LATTICE2010 } (2010)  121.
  
\bibitem{deDivitiis:2011eh}
  G.~M.~de Divitiis {\it et al.},
  arXiv:1110.6294 [hep-lat].
  
\bibitem{Frezzotti:2000nk}
  R.~Frezzotti {\it et al.} [ Alpha Collaboration ],
  JHEP {\bf 0108 } (2001)  058.
  [hep-lat/0101001].
  
\bibitem{Blossier:2010cr}
  B.~Blossier {\it et al.} [ ETM Collaboration ],
  Phys.\ Rev.\  {\bf D82 } (2010)  114513.
  [arXiv:1010.3659 [hep-lat]].

\bibitem{Colangelo:2010et}
  G.~Colangelo {\it et al.},
  Eur.\ Phys.\ J.\  {\bf C71 } (2011)  1695.
  [arXiv:1011.4408 [hep-lat]].
  
\bibitem{Gasser:2003hk}
  J.~Gasser, A.~Rusetsky, I.~Scimemi,
  Eur.\ Phys.\ J.\  {\bf C32 } (2003)  97-114.
  [hep-ph/0305260].

\bibitem{Alexandrou:2008tn}
  C.~Alexandrou {\it et al.} [ European Twisted Mass Collaboration ],
  Phys.\ Rev.\  {\bf D78 } (2008)  014509.
  
\bibitem{Gasser:1982ap}
  J.~Gasser, H.~Leutwyler,
  Phys.\ Rept.\  {\bf 87 } (1982)  77-169.

\end{thebibliography}
\end{document}